\documentclass[twocolumn,showpacs]{revtex4}
\usepackage[pctex32]{graphicx}

\begin{document}

\title{Application of detrended fluctuation analysis to monthly
average of the maximum daily temperatures to resolve different
climates}

\author{M.L. Kurnaz}
\affiliation{Department of Physics, Bogazici University, 34342
Bebek Istanbul}

\begin{abstract}

Detrended fluctuation analysis is used to investigate correlations 
between the monthly average of the maximum daily temperatures for 
different locations in the continental US and the different 
climates these locations have. When we plot the scaling exponents 
obtained from the detrended fluctuation analysis versus the 
standard deviation of the temperature fluctuations we observe 
crowding of data points belonging to the same climates. Thus, 
we conclude that by observing the long-time trends in the 
fluctuations of temperature it would be possible to distinguish 
between different climates.
\end{abstract}

\maketitle

\section{INTRODUCTION\protect\\ }
\label{sec:level1}

Ever since it was first realized that humanity is capable of 
changing its natural surroundings more rapidly than nature can 
repair it, model-based predictions about climate change has been 
one of the main research areas in science. While many 
climatologists firmly believe that there is strong evidence 
for our interference in nature in general and in climate in 
particular, a few others disagree. 

Before we can even attempt to settle this debate, there are 
quite a few questions that we need to answer about the climate 
of the past. Nature provides us with different forms of data 
which we can convert to climatic data. For datasets reflecting 
the near past, it is rather simple to obtain information about 
the climate as we can always rely on the recorded history. For 
example, tree ring history from Scandinavia indicates that there 
was a period of high temperatures between 9th and 13th centuries, 
called the "Medieval Warm Period" \cite{Cook13050}. The tree 
ring temperature history for this period agrees with the glacier 
data, and has been supported by the historical records of Norse 
seafaring and colonization around the North Atlantic at the end 
of the 9th century. It is known that during this time the warmer 
climate helped in producing greater harvests in Iceland and parts 
of Greenland \cite{Bryson13060} which in turn helped the colonies 
in Greenland. But when we go to older times, it becomes more and 
more difficult to relate the paleoclimatic proxies to the climate 
of a certain region. From paleoclimatic proxies, like tree ring 
indices or stacked oxygen-isotope curves derived from deep sea 
cores, we normally obtain fluctuations of the local temperatures 
rather than their absolute values. Therefore, it is difficult to 
construct a method to characterize different climates based 
solely on the fluctuations of temperature values. 

The weather forecast to first approximation, is a rather simple 
issue. A cold day is usually followed by a cold day, and a warm 
day is usually followed by a warm day. On a larger scale, a colder 
week is usually followed by a warmer week which corresponds to 
the average duration of the general weather regimes. But as the 
longer timescales are governed by different processes like 
circulation patterns and sometimes even influenced by trends like 
global warming, defining long-term correlations becomes more 
difficult. 

In order to separate the trends and the correlations we 
need to eliminate the trends in our temperature data. Several 
methods are used effectively for this purpose: rescaled range 
analysis (R/S) \cite{Mandelbrot12180, Mandelbrot12190, Bodri13500,
Bodri13530}, wavelet techniques (WT) \cite{Arneodo5250, 
Koscielny-Bunde13240} and detrended fluctuation analysis (DFA) 
\cite{Peng13540, Koscielny-Bunde13250}.

Analysis of the temperature fluctuations over a period of decades 
on different parts of the globe has already showed the 
effectiveness of the application of detrended fluctuation 
analysis to characterize the persistence of weather and climate 
regimes. DFA and WT have been applied to study temperature and 
precipitation correlations in different climatic zones on the 
continents and also in the sea surface temperature of the oceans. 
The recent results show that the temperatures are long range 
power-law correlated. The long-term persistence of the 
temperatures can be characterized by an auto-correlation function 
$C(n)$ of temperature variations where $n$ is the time between 
the observations. The auto-correlation function decays as 
$C(n)\sim n^{-\gamma}$. Even though there is some disagreement 
on the value of the exponent $\gamma$, the fact that the 
persistence of the temperatures can be characterized by this 
auto-correlation function is firmly established. Different 
groups have used R/S, DFA and WT analysis and have shown that 
this exponent $\gamma$  has roughly the same value 
$\gamma \simeq 0.7$ for the continental stations \cite{Bodri13500, 
Bodri13530, Koscielny-Bunde13250, Bunde13180, Kantelhardt13220, 
Eichner13120, Weber2830, Talkner3550}.  The exponent $\gamma$ 
is found to be roughly 0.4 for island stations \cite{Bunde13180, 
Eichner13120} and sea surface temperature on the oceans 
\cite{Bunde13180, Monetti13140}. This method has also been applied 
to the temperature predictions of coupled atmosphere-ocean general 
circulation models \cite{Bunde100, Govindan13190, Blender13260, 
Fraedrich13270, Fraedrich12720} but there is disagreement on the 
actual value of the exponent $\gamma$. On one side it is argued 
that the exponent does not change with the distance from the oceans 
\cite{Bunde100, Eichner13120} and is roughly $\gamma \simeq 0.7$. 
On the other side it is said that the scaling exponent is roughly 
1 over the oceans, roughly 0.5 over the inner continents and about 
0.65 in transition regions \cite{Blender13260, Fraedrich13270, 
Fraedrich12720}.

Previous work in this area also shows that there is a slight 
variation in the scaling exponent between the low-elevation, 
mountain, continental and maritime stations \cite{Weber2830,
Talkner3550}. Even though these variations are between  
$\gamma \simeq 0.5$ and $\gamma \simeq 0.7$ the fact that they 
show a correlation with location and elevation indicates that a 
relationship between the statistical nature of the temperature 
fluctuations and the climate can be established.

\section{METHOD\protect\\ }
\label{sec:level2} 

Daily temperature data have a non-stationary nature due to 
seasonal trends. To remove this seasonal trend effect, the mean 
temperature for each day over all the years in the data set is 
determined. Subsequently, mean daily maximum temperature was 
obtained by calculating the fluctuation from the mean daily 
maximum temperature,

\begin{equation}
\Delta T_{i} = T_{i} - <T_{i}>
\end{equation}

where $<T_{i}>$ is the mean daily maximum temperature. Similarly 
we can also use the mean daily average temperature or the mean 
daily minimum temperature, and the use of average or minimum 
temperature instead of maximum temperatures does not change the 
outcome of the analysis \cite{Talkner3550}. To remove the remaining 
linear trends (the average temperature for some years can be 
higher or lower than the average temperature of the time series 
for that location as a result of atmospheric processes), we 
applied DFA \cite{Peng13540}, which essentially enables us to 
investigate long-term correlations in the data by getting rid 
of the trends.

The standard calculation of the autocorrelation function is 
hindered by the noise and nonstationarity in the data. Instead 
of calculating $C(n)$ directly and reducing the noise in the time 
series, a running sum of the temperature fluctuations is considered,

\begin{equation}
y (m) = \sum_{i=1}^n \Delta T_{i}
\end{equation}

where $m = 1, … , N$. $N$ is the length of the time series in 
question. The fluctuations   in this sum are related to $C(n)$ by

\begin{equation}
\overline{F(n)^2} \sim n^{2 \alpha}
\end{equation}

Next, the time series of the $y(m)$ is divided into non-overlapping 
intervals of equal length $n$. In each interval, we fit $y(m)$ to a 
straight line ($x(m) = km + d$ for each segment) and calculate the 
detrended square variability $\overline{F(n)^2}$ as

\begin{equation}
\overline{F^{2}(n)} = < \frac{1}{n} \sum_{m = kn + 1}^{(k + 1)n} 
(y(m) - x(m))^{2}>
\end{equation}

with

\begin{equation}
k = 0, 1, 2, . . . , ( \frac{N}{n} - 1).
\end{equation}

If the temperature fluctuations were uncorrelated (i.e. white noise) 
we would expect

\begin{equation}
F(n) \sim n^\alpha
\end{equation}

with $\alpha = \frac{1}{2}$. If $\alpha > \frac{1}{2}$, we expect 
long-range power law correlations in the data for the range of 
values considered.

Figure 1 shows $F(n)$ for one US station as an arbitrary example 
(all other stations in our dataset, where daily temperature data 
is available, show similar behavior). For this station, (Millinocket, 
Maine) we have plotted the $F(n)$ for both daily and monthly 
temperature data. To calculate the monthly temperature data we took 
the average of the maximum daily temperatures for each month and used 
DFA on the monthly averages. Figure 1 shows that for a time period 
longer than about 60 days, both the daily and monthly temperature data 
give us the same scaling exponent $\alpha$. This behavior has been 
observed before for daily temperature data \cite{Weber2830,
Talkner3550}. As the results of our analysis for daily and monthly 
data at all the sampled locations agrees for long time scales 
($n >$ 60 days) the results presented in this paper have been 
obtained from the analysis of monthly averages. Both the daily 
($n >$ 60 days) and monthly averages span about two decades, 
indicating correlations of 30 years or more. We are aware of the 
fact that by taking the monthly averages we are effectively losing 
information about the dataset, i.e. about correlations at a short 
time scale. However, as our main interest is in finding correlations 
in the longer time scales, in the light of the information obtained 
from Figure 1, we believe that this loss does not change the behavior 
of these data at longer time scales. In addition, much wider and 
reliable availability of monthly averages lead us to sacrifice short 
time scales for the sake of obtaining longer time series of data.

\begin{figure}[!]
\includegraphics[bb= 0 0 241 179]
{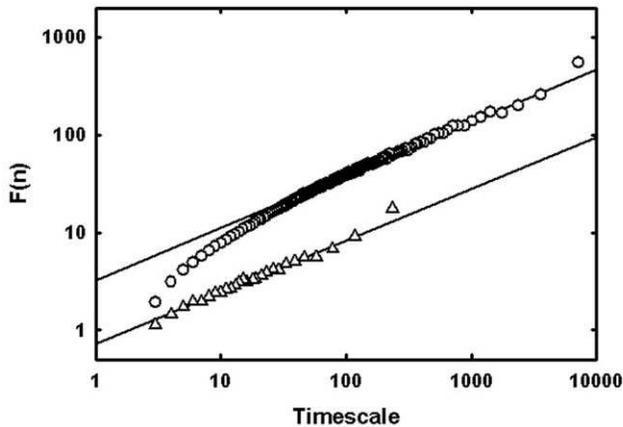}
\caption{Detrended variability $F(n)$ is plotted against a timescale. 
The circles represent daily maximum temperatures, the triangles 
represent monthly average temperatures. The slope of both of the 
curves gives $\alpha = 0.53 \pm 0.03$ (the slope for the daily maximum 
temperatures is calculated for the time period longer than 60 days). 
Data belongs to station, Millinocket, Maine.} 
\label{fig1}
\end{figure}

In the present work we have investigated temperature fluctuations for 
129 weather stations in the continental US. The data has been obtained 
from the U.S. Historical Climatology Network \cite{US13720}. From 
the available data we have chosen the stations with the longest 
records. We did not include in our analysis datasets with data 
shorter than 75 years, and the longest dataset we had spanned 110 
years at several locations in the dataset (all ending in 1994). The 
data come from the coastal regions of California (14 stations), 
Alabama (15 stations), Maine (12 stations), New Mexico (24 stations), 
West Virginia (13 stations), Michigan (17 stations), Montana (14 
stations), and Arizona (20 stations). 

For these 129 continental U. S. stations the value of the exponent 
is found to be $\alpha = 0.60 \pm 0.05$. The individual error bars 
for the data points have been of the order of $\Delta \alpha 
\simeq 0.01$. Figure 2 gives a summary of the scaling exponents 
obtained from these 129 stations. As we can see from this figure, 
consistent with the earlier observations \cite{Eichner13120, 
Govindan13190, Koscielny-Bunde13240, Koscielny-Bunde13250} we 
obtain scaling exponents in the range of 0.52 to 0.72. The scaling 
exponents for the eight different states are coastal California 
($\alpha = 0.63 \pm 0.06$), Alabama 
($\alpha = 0.56 \pm 0.03$), Maine ($\alpha = 0.60 \pm 0.04$), 
West Virginia ($\alpha = 0.56 \pm 0.02$), New Mexico 
($\alpha = 0.64 \pm 0.04$), Arizona ($\alpha = 0.61 \pm 0.03$), 
Michigan ($\alpha = 0.58 \pm 0.02$), and Montana 
($\alpha = 0.58 \pm 0.02$) respectively. 

\begin{figure}[!]
\includegraphics[bb= 0 0 241 187]{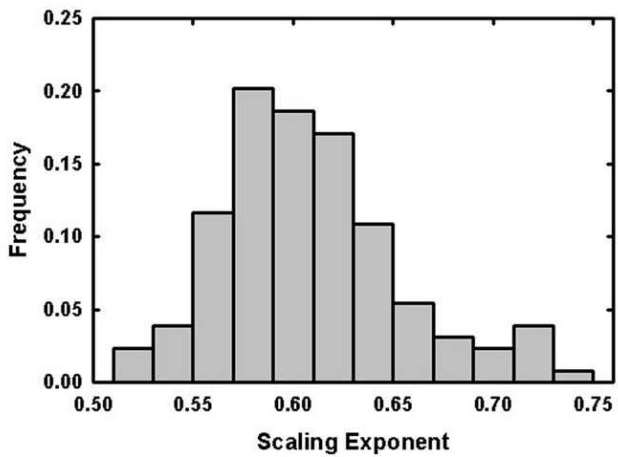}
\caption{The histogram of the scaling exponents. The average from 
the 120 station gives a scaling exponent of $\alpha = 0.60 \pm 0.05$.} 
\label{fig2}
\end{figure}

It has been suggested that the value of the exponent $\alpha$ 
depends on the geographic location (distance from the oceans 
\cite{Fraedrich13270} and elevation \cite{Weber2830}). To 
investigate this effect of location on the exponent $\alpha$ 
we first looked at the correlations between the elevation of the 
weather stations and the resulting exponent $\alpha$ in Figure 3. 
It is very difficult to observe any trends in the data with 
changing elevation as most of the correlations are within one standard 
deviation. We might say that at elevations of about 200 meters 
(the stations at 100-250 meters give an exponent $\alpha = 0.57 \pm 0.03$, 
the scaling exponent is slightly lower than the coastline (the stations 
at 0-100 meters give an exponent $\alpha = 0.62 \pm 0.05$). Above 
the elevation of 250 meters the scaling exponent increases slightly 
with elevation. 

\begin{figure}[!]
\includegraphics[bb= 0 0 241 181]{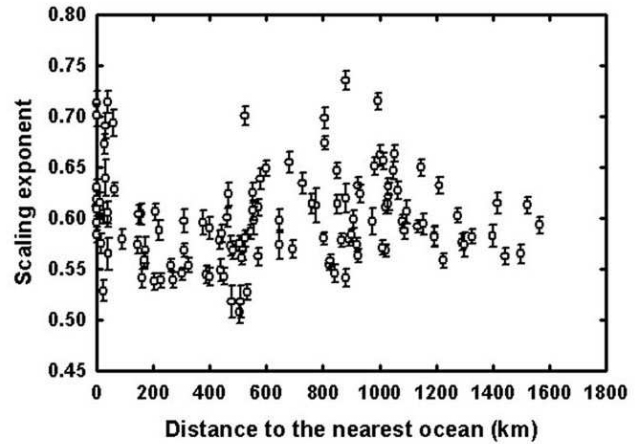}
\caption{The dependence of the scaling exponent on the elevation of the 
weather station.} 
\label{fig3}
\end{figure}

Previous work suggested that the scaling exponent changed from 
1 over the oceans to 0.5 over the inner continents. The coastal 
regions appeared as transition regions corresponding to a scaling 
exponent of about 0.65 \cite{Fraedrich13270}. Figure 4 shows the 
relation between the distance between the station and nearest ocean, 
and the scaling exponent. We observe no change in the scaling exponent 
for the inner continental stations, which contradicts the previous work, 
but the distance of the observational station previously used, was 
about 2000 kms from any ocean, whereas in our data set, the largest 
distance between any of the stations and the nearest ocean is about 
1600 kms. Therefore we cannot comment on the possibility that the 
exponent changes to smaller values for even more "inner continental" 
stations. Latest work also agrees with our data in showing no change 
with increasing distance from the coastline \cite{Eichner13120}.

\begin{figure}[!]
\includegraphics[bb= 0 0 241 176]{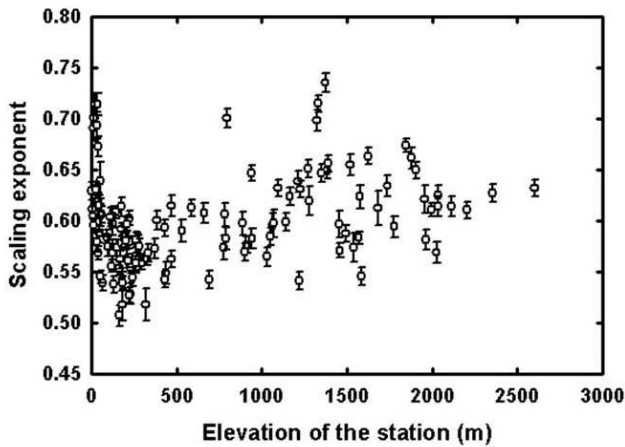}
\caption{The dependence of the scaling exponent on the distance 
of the weather station from the nearest coastline (ocean).} 
\label{fig4}
\end{figure}

When we look at the scaling exponents and consider only the geographic 
locations of the stations it is almost impossible to distinguish these 
states from each other unambiguously.  However a striking result can be 
observed if we plot the standard deviation of the temperature 
fluctuations versus the scaling exponent observed from that station. 
These standard deviations of the temperature fluctuations are calculated 
over the whole dataset where each point consists of a monthly average of 
daily maximum temperatures. It is known that the standard deviation is a 
correct measure of fluctuations only if the underlying distribution is a 
pure Gaussian, and daily and monthly temperatures are known to show skewed 
distributions. The coefficient of skewness is defined as \cite{Press13730}

\begin{equation}
Skewness (x_{1} . . . x_{N}) =  \frac{1}{n} \sum_{j = 1}^{N} 
[\frac{x_{j}-\overline{x}}{\sigma}]^{3}
\end{equation}

where $\sigma = \sigma (x_{1} . . . x_{N})$ is the distribution's standard 
deviation. A positive value of skewness signifies a distribution where an 
asymmetric tail extends out towards more positive $x$ values and a negative 
value of skewness gives a distribution with an asymmetric tail extending out 
towards more negative $x$ values. For the idealized case of a Gaussian 
distribution, the standard deviation of the coefficient of skewness is 
approximately $\sqrt{6/N}$. In real life it is a good practice to account 
for skewnesses only if the coefficient of skewness obtained from the 
distribution is many times larger than this value \cite{Press13730}. We 
have randomly chosen five of the stations to test for skewness. For all 
of these stations the period of the time series was 102 years. For 
comparison the standard deviation of the coefficient of skewness for a 
Gaussian distribution is $\sqrt{6/1224}=0.07$. The coefficients of skewness 
for these stations are 0.03 (Brewton, AL), 0.09 (Berkeley, CA), -0.05 
(Crow Agency, MT), -0.04 (State University, NM), and -0.06 (Glenville, WV). 
Figure 5 shows the frequency of the deviations from the monthly average 
temperatures for a sample station, Glenville, WV. As a side note, we have 
performed the same skewness analysis for these stations using the average 
monthly maximum temperatures instead of the deviation from the average 
maximum temperatures. The skewness of this data was significantly higher 
justifying the point that climatic data itself can have a strongly skewed 
distribution (-0.17 (Brewton, AL), -0.28 (Berkeley, CA), -0.13 (Crow Agency, 
MT), -0.17 (State University, NM), and -0.20 (Glenville, WV)). Therefore 
we decided that the standard deviation of temperature fluctuations from the 
average temperatures can be used in statistical description of a dataset as 
long as the climate does not change during that given period of time.   

\begin{figure}[!]
\includegraphics[bb= 0 0 241 177]{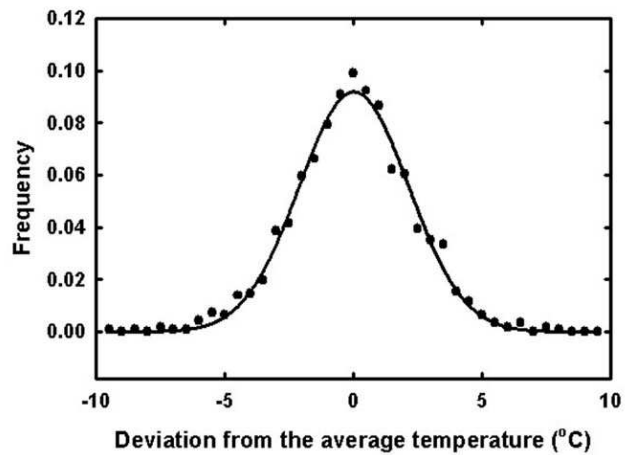}
\caption{The frequency of the deviation of the monthly temperatures 
from the monthly average temperatures. The fitting line is to a 
Gaussian distribution.} 
\label{fig5}
\end{figure}

We are aware that state or country boundaries are not good indicators for 
identifying different regions, however, within many small states or 
countries, the climate does not change significantly. In cases where the 
climate is different in different parts of the state, we have either 
ignored the data from those stations, or just analyzed one part of the 
state as in the case of coastal California. Therefore, we can safely 
assume that the climates in these states can be classified as Humid 
Subtropical - Mediterranean (Coastal California), Humid Subtropical - 
East Coasts (Alabama), Humid Continental - Hot Summers - Year around 
precipitation (West Virginia), Humid Continental - Mild Summers - Year 
around precipitation (Maine) and Dry/Arid - Hot - Low Latitude desert 
(New Mexico).

Figures 6-10 give the scaling exponent for coastal California, Alabama, 
Maine, West Virginia, and New Mexico versus the standard deviation of the 
temperature fluctuations, respectively. In these figures we can identify 
that the scaling exponents crowd different regions of the graph indicating 
a possibility that these different climates can be distinguished from each 
other using the method described above. We must be clear about one point: 
If we have only the standard deviations of the temperature fluctuations and 
the scaling exponents resulting from those distributions, obtaining clusters 
which would indicate different climates, is extremely difficult. However the 
question we are trying to answer is simpler: We know where the stations are 
located (on the standard deviation of temperature fluctuations versus the 
scaling exponents map) for Humid Subtropical - Mediterranean climate. Can we 
now identify an unknown station as having a Humid Subtropical - Mediterranean 
climate? To be able to answer this question we have used the support vector 
machine (SVM) algorithm \cite{Vapnik13740} for data classification. 

\begin{figure}[!]
\includegraphics[bb= 0 0 241 172]{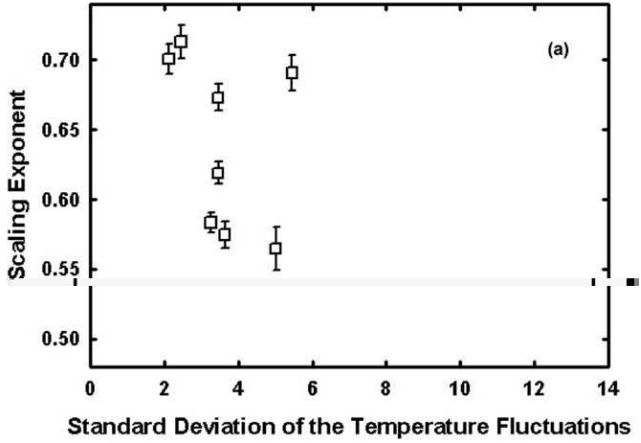}
\caption{The scaling exponents plotted against the standard deviation of 
the temperature fluctuations for coastal California. } 
\label{fig6a}
\end{figure}

\begin{figure}[!]
\includegraphics[bb= 0 0 241 172]{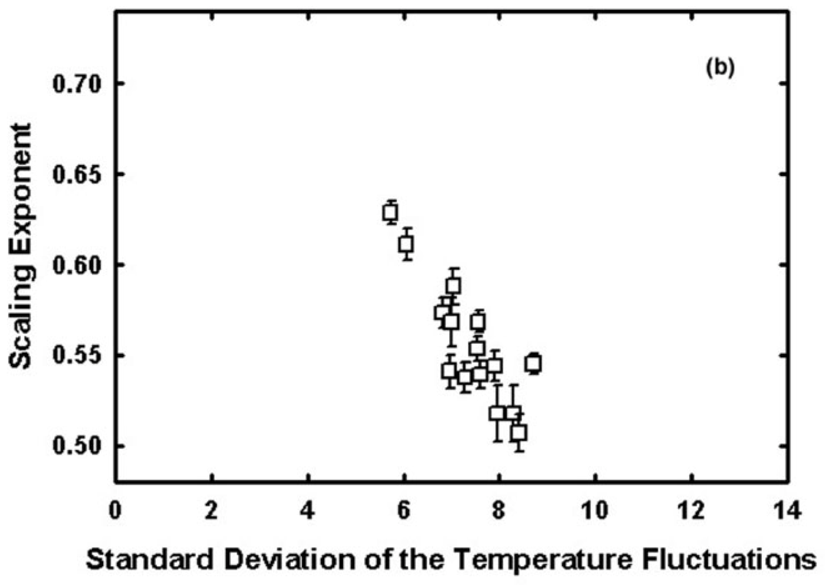}
\caption{The scaling exponents plotted against the standard deviation of 
the temperature fluctuations for Alabama. } 
\label{fig6b}
\end{figure}

\begin{figure}[!]
\includegraphics[bb= 0 0 241 170]{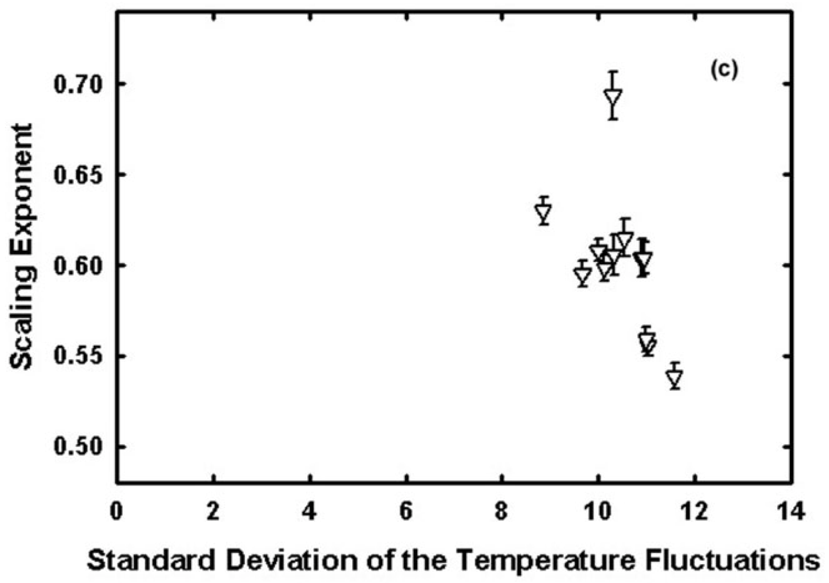}
\caption{The scaling exponents plotted against the standard deviation of 
the temperature fluctuations for Maine. } 
\label{fig6c}
\end{figure}

\begin{figure}[!]
\includegraphics[bb= 0 0 241 174]{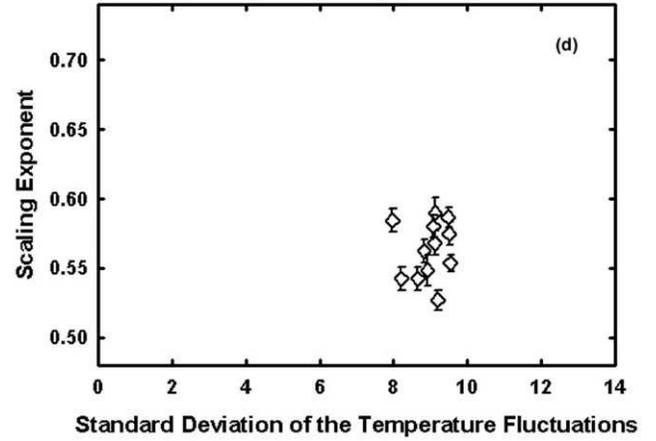}
\caption{The scaling exponents plotted against the standard deviation of 
the temperature fluctuations for West Virginia. } 
\label{fig6d}
\end{figure}

\begin{figure}[!]
\includegraphics[bb= 0 0 241 173]{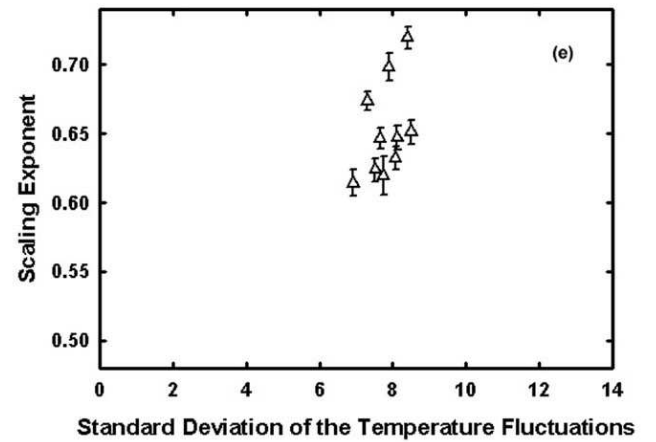}
\caption{The scaling exponents plotted against the standard deviation of 
the temperature fluctuations for New Mexico. } 
\label{fig6e}
\end{figure}

A data classification task normally uses training and testing data. Each 
instance in the training data set consists of one target value (in our case 
belonging to a specific climate type) and several features (like the standard 
deviation of the temperature fluctuations and the scaling exponent). The aim 
of SVM \cite{Joachims13750} is to produce a model which then predicts the 
target value of data instances in the testing set. In training set we have 
used 10 positive target values (belonging to the climate class we are 
analyzing) and 20 negative target points (belonging to different climate 
classes). In the testing set, we supplied the algorithm with all of the 
stations in our data set and asked the program to identify the different 
regions. The performance of such an algorithm is usually quantified by its 
accuracy during the test phase which mainly depends on the correct treatment 
of true positives (TP) and true negatives (TN). It is usually also important 
to distinguish between two types of errors: A false positive (FP) and a false 
negative (FN). Consequently, the performance of the prediction is better 
judged if we add two more quantifiers, sensitivity and specificity. The 
accuracy of the data classification is defined as the ratio between the 
number of correctly identified samples and the total number of samples:

\begin{equation}
accuracy = \frac{TP+TN}{TP+FP+TN+FN} 
\end{equation}

The sensitivity is the ratio between the number of true positive 
predictions and the number of positive instances in the test set:

\begin{equation}
sensitivity = \frac{TP}{TP+FN} 
\end{equation}

Finally, the specificity is defined as the ratio between the number 
of true negative predictions and the number of negative instances 
in the test set:

\begin{equation}
specificity = \frac{TN}{FP+TN} 
\end{equation}

For our initial analysis, we used only the five different climates 
mentioned above. The results are summarized in Table 1. Considering 
that we have a test set of 78 stations the deviations from $100\%$ for 
any of the accuracy, sensitivity, and specificity values are caused 
by at most 4 stations identified either as false positives or false 
negatives. We believe that this small discrepancy is caused by 
microclimatic behavior for those stations. For example, the only 
false positive for Humid Subtropical - Mediterranean climate comes 
from the only coastal station in Alabama.

\begin{center}
\begin{table*}[!]
\caption{SVM analysis of the 78 weather stations in 5 different 
climate zones. Accuracy, sensitivity, and specificity are defined 
in the text.} 
\label{sums}
\begin{tabular}{|l|c|c|c|}
\hline Climate Type & Accuracy & Sensitivity & Specificity \\ 
\hline Humid Subtropical - Mediterranean (Coastal California) & 
$98.7 \%$ & $100.0 \%$ & $98.4 \%$ \\ 
\hline Humid Subtropical - East Coasts (Alabama) &
$96.2 \%$ & $100.0 \%$ & $95.2 \%$ \\ 
\hline Humid Continental - Mild Summers - Year around precipitation 
(Maine) & $98.7 \%$ & $91.7 \%$ & $100.0 \%$ \\ 
\hline Dry/Arid - Hot - Low Latitude desert (New Mexico) &
$96.2 \%$ & $100.0 \%$ & $94.4 \%$ \\ 
\hline Humid Continental - Hot Summers - Year around precipitation 
(West Virginia) & $94.9 \%$ & $100.0 \%$ & $93.9 \%$ \\ 
\hline
\end{tabular}%
\end{table*}
\end{center}

To test this method even further, we have used SVM algorithm on the 
remaining 51 stations. Out of these 51 stations, 31 were in Montana 
and Michigan (Humid Continental - Mild Summers - Year around 
precipitation climate like Maine) and 20 were in Arizona (Dry/Arid - 
Hot - Low Latitude desert climate like New Mexico). Montana and Michigan 
are geographically different from Maine both in distance from the 
coastline and elevation. Arizona is geographically similar to New 
Mexico in distance from the coastline and elevation. In the analysis, 
we have treated Maine as our known climate, and together with the other 
states, Montana and Michigan as our "unknowns". Table 2 shows the 
results of this analysis. As expected, accuracy, sensitivity and 
specificity dropped slightly when we considered significantly different 
geographic locations. However, we still have more than 95
predicting the climate of Michigan and Montana based solely on this 
analysis. In the second part of this analysis, we have considered New 
Mexico as our known climate, and together with the other states in our 
dataset, Arizona as our "known" climate. The results show that all of 
the stations in Arizona are correctly identified as belonging to 
Dry/Arid - Hot - Low Latitude desert climate except for one station. 
Improvement in the statistics is caused by increasing the number of 
stations.

\begin{center}
\begin{table*}[!]
\caption{SVM analysis of the whole data set in 2 different climate 
zones. For Humid Continental climate, the weather stations in Maine 
are taken as "known" and data classification is performed to see 
whether the weather stations in Montana and Michigan are picked out 
as well for this climate. Similar analysis is done for the New 
Mexico and Arizona pair.} 
\label{sums}
\begin{tabular}{|l|c|c|c|}
\hline Climate Type & Accuracy & Sensitivity & Specificity \\ 
\hline Humid Continental - Mild Summers - Year around precipitation 
(Maine) & $95.4 \%$ & $90.7 \%$ & $97.7 \%$ \\ 
\hline Dry/Arid - Hot - Low Latitude desert (New Mexico) &
$96.9 \%$ & $97.7 \%$ & $96.5 \%$ \\ 
\hline
\end{tabular}%
\end{table*}
\end{center}

\section{CONCLUSION\protect\\ }
\label{sec:level3}

In this study, the variability of the weather in different parts of 
the continental US, as an example of different climates, has been 
investigated. Our results suggest that different climates can be 
readily distinguished using the detrended fluctuation analysis 
method on the fluctuations of the maximum daily temperatures. Even 
though we have used state boundaries to define the climates, as 
long as a mild, subtropical, Mediterranean climate exists in Coastal 
California, this method should be equally applicable to distinguish 
this climate from New Mexico where a dry, arid, hot desert climate 
is observed. 

The results presented here are preliminary, based on stations with 
known climates. The real challenge lays in the future ability of this 
method to be applied to paleoclimatic data to reveal structure over 
timescales not only of the order of decades but that of millions of 
years. This point stays speculative as no reliable monthly data exists 
beyond 218 years \cite{Koscielny-Bunde13240} (to our knowledge). 
However, the fact that this data does not seem to scatter appreciably 
at longer timescales \cite{Monetti13140} gives us hope about expanding 
the use of the method mentioned above.

\begin{acknowledgements}
Data were provided by U.S. Historical Climatology Network and the National 
Climatic Data Center.
\end{acknowledgements}

\end{document}